\documentclass[twocolumn,amsmath,prl,superscriptaddress,nofootinbib]{revtex4-2}

\usepackage{stix}
\usepackage{upgreek}
\usepackage{bm}
\usepackage{pifont}
\usepackage{graphicx}
\usepackage{color}
\usepackage[colorlinks=true, pdfstartview=FitV, linkcolor=blue, citecolor=blue, urlcolor=blue]{hyperref}
\usepackage{ulem}

\newcommand{\Red}[1]{{#1}}

\newcommand{\INH}{\sigma_{\mathrm{INH}}}
\newcommand{\BCD}{\sigma_{\mathrm{BCD}}}
\newcommand{\Drude}{\sigma_{\mathrm{Drude}}}
\newcommand{\mathd}{\mathrm{d}}
\newcommand{\tick}{\ding{51}}
\newcommand{\cross}{\ding{55}}

\begin{document}

\title{Intrinsic nonlinear Hall effect in antiferromagnetic tetragonal CuMnAs}

\author{Chong \surname{Wang}}
\affiliation{Department of Physics, Carnegie Mellon University, Pittsburgh, Pennsylvania 15213, USA}

\author{Yang \surname{Gao}}
\email{ygao87@ustc.edu.cn}
\affiliation{ICQD, Hefei National Laboratory for Physical Sciences at Microscale, University of Science and Technology of China, Hefei, Anhui 230026, China}
\affiliation{Department of Physics, University of Science and Technology of China, Hefei, Anhui 230026, China}

\author{Di \surname{Xiao}}
\email{dixiao@uw.edu}
\affiliation{Department of Physics, Carnegie Mellon University, Pittsburgh, Pennsylvania 15213, USA}

\begin{abstract}
\Red{Detecting the orientation of the N\'eel vector is a major research topic in antiferromagnetic spintronics. Here we recognize the intrinsic nonlinear Hall effect, which is independent of the relaxation time, as a prominent contribution to the time-reversal-odd second order conductivity and can be used to detect the flipping of the N\'eel vector. In contrast, the Berry-curvature-dipole-induced nonlinear Hall effect depends linear on relaxation time and is time-reversal-even.} We study the intrinsic nonlinear Hall effect in an antiferromagnetic metal: tetragonal CuMnAs, and show that its nonlinear Hall conductivity can reach the order of mA/V$^2$. The dependence on the chemical potential of such nonlinear Hall conductivity can be qualitatively explained by a tilted massive Dirac model. Moreover, we demonstrate its strong temperature dependence and briefly discuss its competition with the second order Drude conductivity. Finally, a complete survey of magnetic point groups are presented, providing guidelines for finding more antiferromagnetic materials with the intrinsic nonlinear Hall effect.
\end{abstract}

\maketitle

\textit{Introduction}.---\Red{Electrically detecting the flipping (180$^\circ$ rotation) of the N\'eel vector is a crucial task in antiferromagnetic spintronics~\cite{jungwirth_antiferromagnetic_2016,Baltz.2018.Tserkovnyak}. Since the N\'eel vector and its flipped image are related by time reversal symmetry, the time-reversal-odd ($T$-odd) part of the conductivity tensor can be utilized to distinguish them. However, many antiferromagnets respects the combination of spatial and time reversal symmetry ($PT$), which forbids the $T$-odd part of the linear conductivity tensor. In contrast, for the second order conductivity tensor, $PT$ symmetry forbids the time-reversal-even ($T$-even) part and allows the existence of the $T$-odd part. Therefore, second order conductivity is an ideal quantity to detect the N\'eel vector flipping in $PT$-symmetric antiferromagnets. In fact, one pioneering experiment detected the N\'eel vector flipping by measuring the sign of the second order conductivity in antiferromagnetic tetragonal CuMnAs~\cite{godinho_electrically_2018}. Despite the success of the experiment, the underlying microscopic mechanism of this second order conductivity has not been fully recognized.

The second order conductivity tensor is defined as the quadratic current response $\bm{J}$ to electric field $\bm{E}$: $J^\alpha = \sum_{\beta,\gamma}\sigma^{\alpha\beta\gamma} E^\beta E^\gamma$ ($\alpha$, $\beta$ and $\gamma$ are Cartesian indices). $\sigma$ can be separated into an Ohmic-type part and a Hall-type part~\cite{tsirkin-SeparationHallOhmic-2021}. The Ohmic part includes a second-order Drude conductivity, which is $T$-odd and quadratically dependent on the relaxation time $\tau$~\cite{holder2020consequences,watanabe2020nonlinear,watanabe_chiral_2021,zelezny-UnidirectionalMagnetoresistanceSpinorbit-2021}. The Hall part also includes a $T$-odd contribution which is independent of the relaxation time, and is therefore called the intrinsic nonlinear Hall effect (INHE)~\cite{gao_field_2014}. The INHE is different from the more well-known Berry curvature dipole (BCD) contribution~\cite{sodemann_quantum_2015,deyo-SemiclassicalTheoryPhotogalvanic-2009,moore-ConfinementInducedBerryPhase-2010,ma2019observation,kang2019nonlinear}. The latter is proportional to $\tau$ and, more importantly, it is a $T$-even quantity and is forbidden in $PT$-symmetric antiferromagnets. The INHE and its thermal counterpart, the intrinsic nonlinear Nernst effect, has been studied in model Hamiltonians such as the loop current model~\cite{gao_orbital_2018}. However, little is known about the INHE in realistic materials.


In this Letter, we show that the INHE is a prominent contribution, and sometimes the dominant contribution, to the $T$-odd second order conductivity.} Using density functional theory (DFT) calculations, we show that $\INH$ in antiferromagnetic tetragonal CuMnAs can reach the order of mA/V$^2$ and is much larger than the Ohmic second order Drude conductivity. We find a large peak accompanied by a sign change of $\INH$ as a function of the chemical potential. This behavior can be explained by a tilted massive Dirac fermion model in which the INHE is shown to have a geometric origin arising from the quantum metric dipole. We find that $\INH$ in tetragonal CuMnAs has a strong temperature dependence and briefly discuss its competition with second order Drude conductivity. Finally, a complete survey of magnetic point groups is presented, providing guidelines for finding other candidate materials where the INHE can be observed. Our results thus establish the INHE as an important transport phenomena in inversion asymmetric magnetic metals, particularly \Red{in the context of probing the N\'eel vector flipping in antiferromagnets}.

\begin{figure}
\centering
\includegraphics[width=0.989\columnwidth]{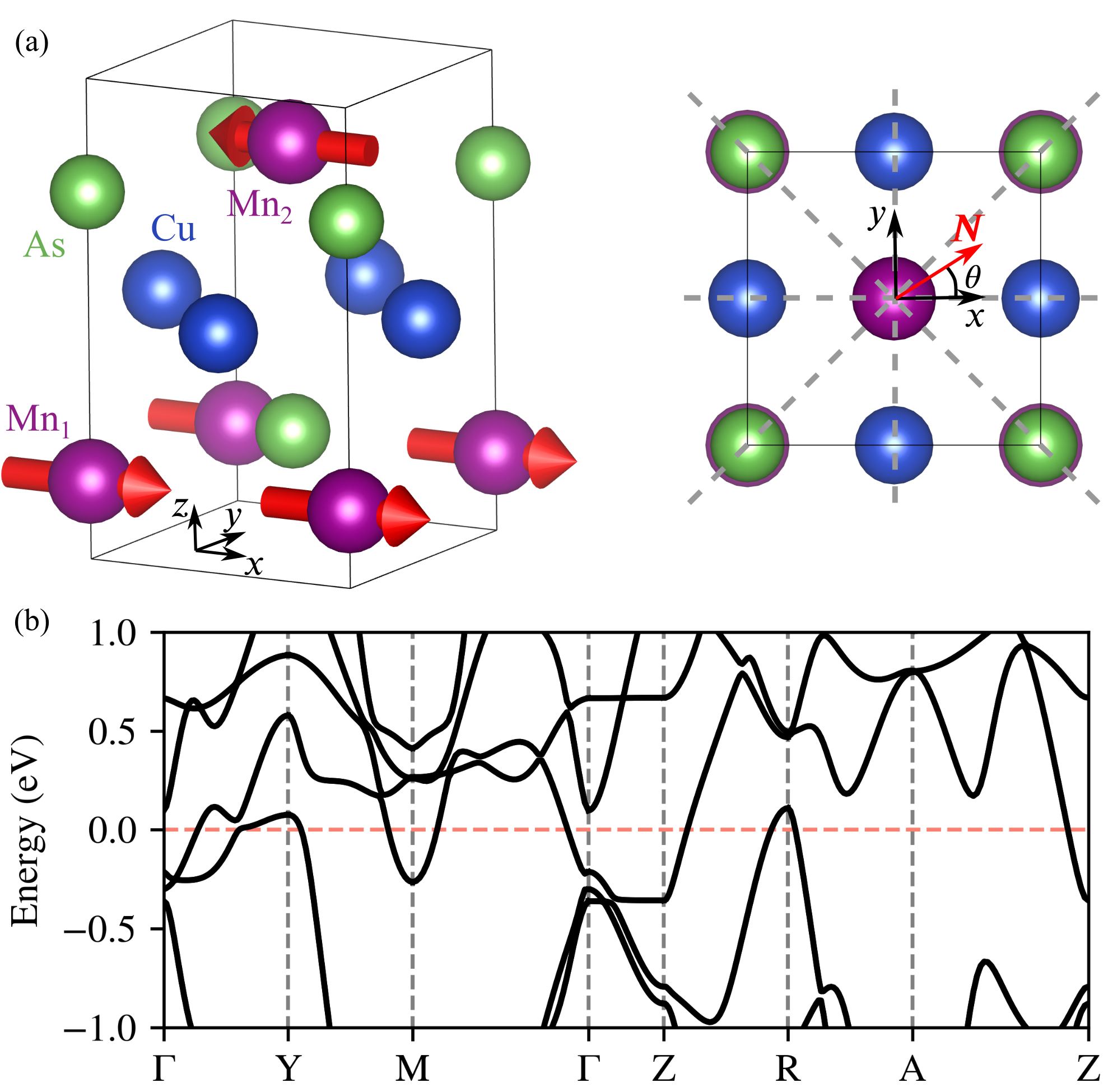}
\caption{(a) The unit cell of tetragonal CuMnAs (left) and its top view (right). Mn$_1$ and Mn$_2$ have opposite magnetic moments, both lying in the $x-y$ plane. Important crystalline symmetries without magnetic order include inversion (inversion center between Mn$_1$ and Mn$_2$ on the left panel), four-fold rotation in the $z$ direction (rotation axis at the Mn atom in the right panel) and four mirror symmetries (mirror planes indicated by dashed gray lines in the right panel). The N\'eel vector $\bm{N}$ is parameterized by the polar angle $\theta$ in the $x-y$ plane. (b) The band structure of CuMnAs for $\theta=0$.}
\label{fig:CuMnAs}
\end{figure}

\textit{Intrinsic nonlinear Hall effect in CuMnAs}.---CuMnAs usually crystallizes in an orthorhombic phase. However, with suitable substrate (e.g., GaAs or GaP), a tetragonal phase can be experimentally grown~\cite{wadley_tetragonal_2013}, which is the focus of this work. Figure~\ref{fig:CuMnAs}(a) presents the atomic structure and important symmetries of tetragonal CuMnAs. Each unit cell contains two Mn planes. The magnetic moment, lying in the Mn plane, is antiferromagnetically ordered between the planes. For magnetic moments lying in different directions within the Mn plane, there is little energy difference. Therefore, the N\'eel vector $\bm{N}$, defined as the difference of the magnetic moments between Mn$_1$ and Mn$_2$ atoms, can be parametrized by the polar angle $\theta$ [see Fig.~\ref{fig:CuMnAs}(a)]. For arbitrary $\theta$, $PT$ symmetry is respected. Electrical manipulation of $\theta$ using current pulses has recently been demonstrated~\cite{wadley_electrical_2016,wadley-CurrentPolaritydependentManipulation-2018}.

The intrinsic nonlinear Hall conductivity can be expressed in terms of band quantities as~\cite{gao_field_2014,gao_orbital_2018}
\begin{equation}
 \INH^{\alpha \beta \gamma} = 2 e^3 \sum_{n,m}^{\epsilon_n \ne \epsilon_m} \mathrm{Re} \int \frac{\mathd^3 \bm{k}}{(2 \uppi)^3} \frac{v^{\alpha}_n A^{\beta}_{nm} A^{\gamma}_{mn} }{\epsilon_n - \epsilon_m} \frac{\partial f (\epsilon_n)}{\partial \epsilon_n} - (\alpha \leftrightarrow \beta),\label{inhexp}
\end{equation}
where $\bm{v}$ is the band velocity, $\epsilon_n$ is the energy of the $n$th Bloch state, $\bm A_{nm} = \langle u_n|\mathrm{i}\nabla_{\bm k}u_m \rangle$ is the Berry connection with $|u_n\rangle$ the periodic part of the $n$th Bloch state, and $e$ is the (positive) elementary charge. $f(\epsilon; T, \mu)$ [$T$ and $\mu$ are omitted in Eq.~(\ref{inhexp})] is the Fermi-Dirac occupation number for energy $\epsilon$ at temperature $T$ and chemical potential $\mu$. For now we work at $T=0$. We note that $\INH^{\alpha\beta\gamma}$ is antisymmetric with respect to $\alpha$ and $\beta$, therefore it describes a Hall-type current~\cite{tsirkin-SeparationHallOhmic-2021}. Due to the derivative of $f$ in Eq.~(\ref{inhexp}), the INHE only depends on quantities around the Fermi surface and is only relevant for metals. As shown in Fig.~\ref{fig:CuMnAs}(b), CuMnAs is indeed a metal.

\begin{figure}
\centering
\includegraphics[width=0.9645\columnwidth]{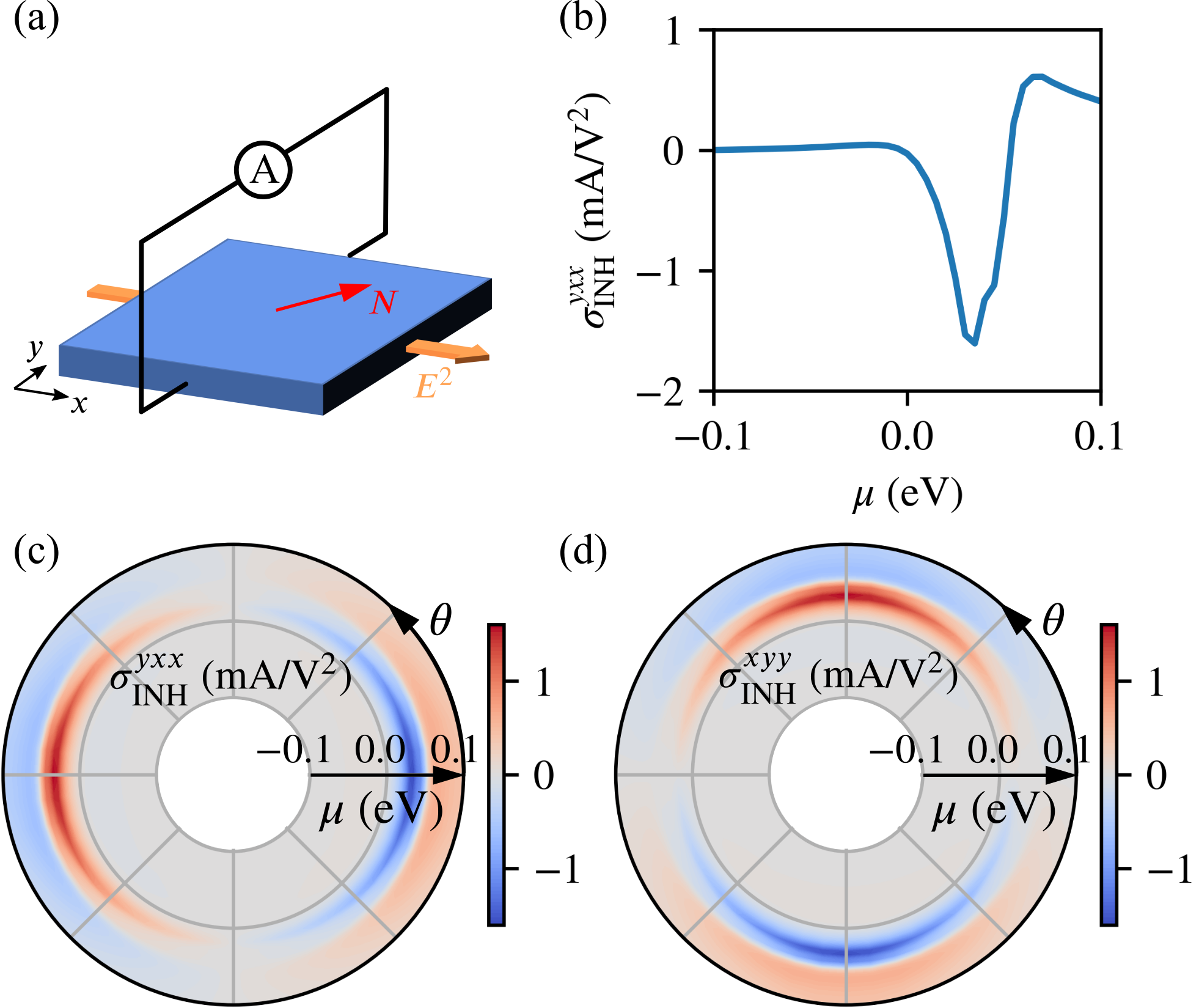}
\caption{\Red{(a) A sketch of the experimental setup to measure $\sigma^{yxx}$. The $x$ and $y$ directions are defined in Fig.~\ref{fig:CuMnAs}. (b) $\INH^{yxx}$ for $\theta=0$ as a function of the chemical potential.  (c-d) Intrinsic nonlinear Hall conductivity $\sigma_{\mathrm{INH}}^{yxx}$ [(c)] and $\sigma_{\mathrm{INH}}^{xyy}$ [(d)] as a function of chemical potential ($\mu$) and orientation of the N\'eel vector $\theta$. $\mu=0$ is the intrinsic chemical potential.}}
\label{fig:INH}
\end{figure}

For a general orientation (not along high symmetric directions such as $\theta=0$) of the N\'eel vector, the magnetic point group of tetragonal CuMnAs is 2$'$/m. The allowed components are $\INH^{xyy}=-\INH^{yxy}$, $\INH^{yxx}=-\INH^{xyx}$, $\INH^{xzz}=-\INH^{zxz}$ and $\INH^{yzz}=-\INH^{zyz}$. Four-fold rotational symmetry along the $z$ axis ($C_{4z}$) demands that $\INH^{xyy (xzz)}(\theta) = \INH^{yxx (yzz)}(\theta + \uppi/2)$. \Red{ Therefore, the only independent components are $\INH^{yxx}$ and $\INH^{yzz}$. Here we focus on $\INH^{yxx}$ and defer the discussion of $\INH^{yzz}$ to the Supplemental Material~\cite{supplemental}.}

We first examine the chemical potential ($\mu$) dependence of $\INH^{yxx}$. A sketch of the experimental setup to measure $\sigma^{yxx}$ is presented in Fig.~\ref{fig:INH}(a). When the N\'eel vector is along the $x$-axis ($\theta = 0$), $\INH^{yxx}$ is generally small for hole doping, but develops a peak and then changes sign rapidly and develops another peak for electron doping [Fig.~\ref{fig:INH}(b)]. This behavior signifies a strong variation of Bloch wave function in the band structure, and will be discussed later. As shown in Fig.~\ref{fig:INH}(b), $\INH^{yxx}$ is on the order of mA/V$^2$.

\Red{Next, we discuss the strong correlation between $\INH$ and the orientation of the N\'eel vector [Fig.~\ref{fig:INH}(c-d)]. In most cases, symmetry arguments can be invoked to explain the correlation. As mentioned in the introduction, time reversal symmetry demands $\INH(\theta) = -\INH(\theta+\uppi)$. Additional constraints on specific components can be derived from crystalline symmetries. For example, when the N\'eel vector is along the $x$ ($y$) direction, $\INH^{xyy}$ ($\INH^{yxx}$) is forbidden by mirror symmetry in the $x$ ($y$) direction. Therefore, the sign of $\INH^{yxx}$ can be used to distinguish the N\'eel vector lying in $\pm x$ direction [Fig.~\ref{fig:INH}(c)], and the sign of $\INH^{xyy}$ can be used to distinguish the N\'eel vector lying in $\pm y$ direction [Fig.~\ref{fig:INH}(d)].}

\begin{figure}
\centering
\includegraphics[width=0.9525\columnwidth]{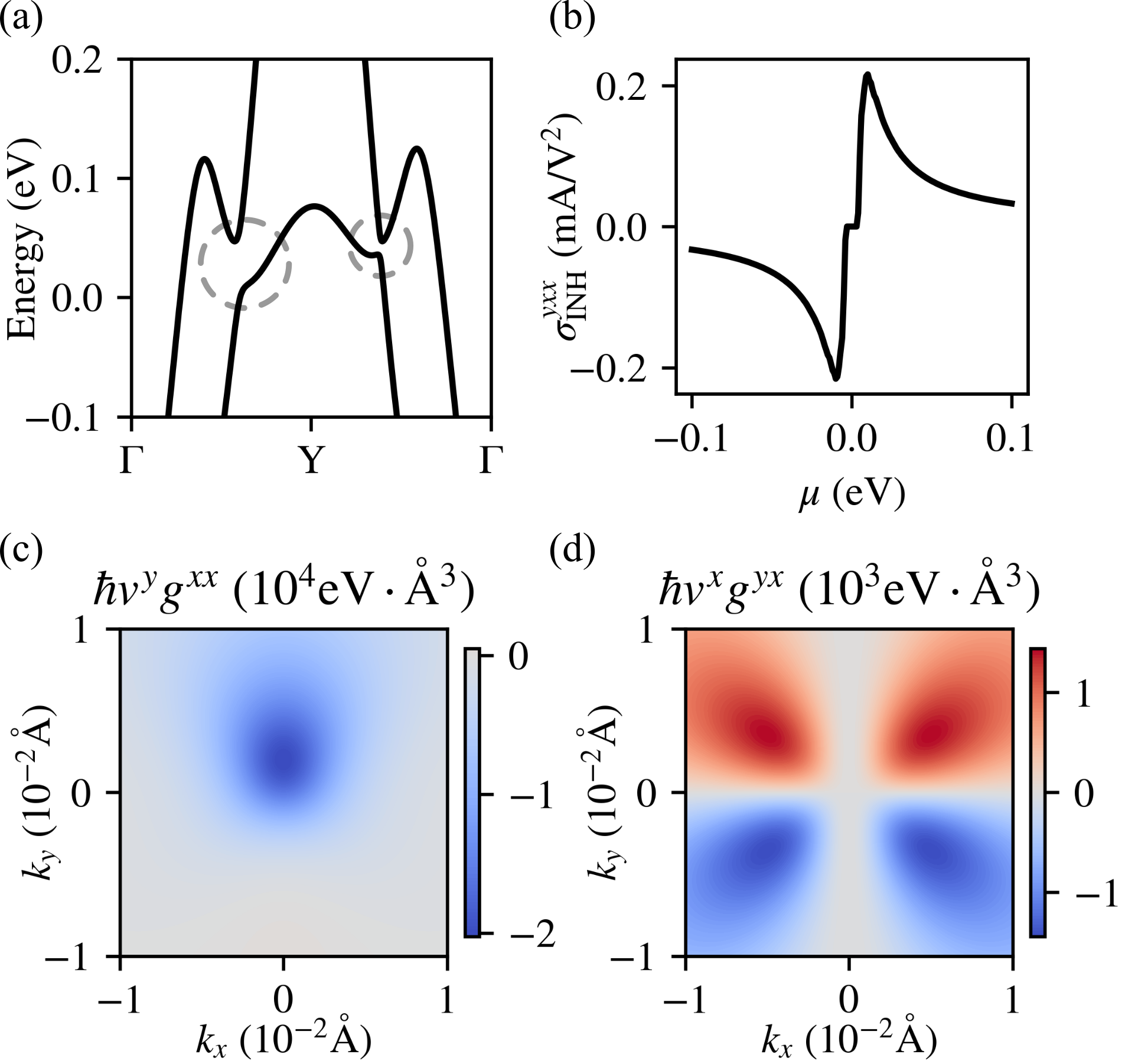}
\caption{(a) shows the band structure of tetragonal CuMnAs along the $\Gamma$-Y-$\Gamma$ line, which contains two avoided crossings. (b) shows the intrinsic nonlinear Hall conductivity of a tilted massive Dirac model. (c) and (d) shows the quantum metric tensor dipole at $k_z=0$ plane for the lower band. $\hbar$ is the reduced Planck constant. Parameters: $t = 0.8$~eV/\AA, $m = 5$~meV.}
\label{fig:Dirac}
\end{figure}

\textit{Tilted massive Dirac point model}.---To gain further insight into the behavior of $\INH^{yxx}$, we analyze the contribution to $\INH^{yxx}$ from different $k$ points at the Fermi surface. We find that the dominant contribution of the peak comes from avoided crossings along the $\Gamma$-X-$\Gamma$ and $\Gamma$-Y-$\Gamma$ lines. Figure~\ref{fig:Dirac}(a) shows the band structure along the $\Gamma$-Y-$\Gamma$ line, where two avoided crossings can be observed. These avoided crossings are common on these high symmetry lines since the little group on these lines have only one allowed representation.

These avoided crossings can be qualitatively modeled
by tilted massive Dirac points. Specifically, we introduce the following Hamiltonian
\begin{equation}
 H_{\mathrm{Dirac}} = k_x \tau_1 \sigma_0 + k_y (\tau_2 \sigma_2 - t \tau_0 \sigma_0) + k_z \tau_3 \sigma_0 + m \tau_2 \sigma_3,\label{Diracham}
\end{equation}
where $\tau$ and $\sigma$ are two sets of Pauli matrices, $t$ is a parameter controlling the tilt of the Dirac point and $m$ controls the gap. $H_{\mathrm{Dirac}}$ respects the $PT$ symmetry, which is represented by $-i\sigma_2 K$, with $K$ the complex conjugate. Therefore, every band is doubly degenerate. 

$\INH$ is expected to have a large value if the chemical potential is near the band edge. As the chemical potential approaches the band edge, the carrier density decreases. However, the Bloch function is rapidly changing, giving rise to a large Berry connection [cf. Eq.~(\ref{inhexp})]. Due to such competition, $\INH$ is zero when the chemical potential is right at the band edge and rapidly develops a peak before decreasing as the chemical potential moves away from the band edge. Indeed, two peaks with opposite signs can be found in $\INH^{yxx}$ as a function of chemical potential [Fig.~\ref{fig:Dirac}(b)], mimicking the sign change in Fig.~\ref{fig:INH}(c). The tilt of the Dirac point is essential for the INHE: without the tilt in the $y$ direction, this model have inversion symmetry, which forbids all components of second order conductivity. This calculation qualitatively explain $\INH^{yxx}$ in CuMnAs and shows that a large INHE should be generally expected for materials with similar band structures.

Using $H_{\mathrm{Dirac}}$, we find that $\INH$ has a geometric origin as it is related to the quantum metric dipole. We use $\nu i$ to label different bands, with $\nu=1,2$ for different sets of bands and $i=1,2$ for $PT$-related degenerate pair. Then Eq.~\eqref{inhexp} becomes~\cite{supplemental}
\begin{equation}
\INH^{\alpha\beta\gamma} = 2 e^3 \sum_{\nu} \int \frac{\mathd^3 \bm{k}}{(2 \uppi)^3} \frac{v_\nu^\alpha g^{\beta\gamma}_\nu}{\epsilon_\nu - \epsilon_{\bar{\nu}}}\frac{\partial f(\epsilon_\nu)}{\partial \epsilon_\nu} - (\alpha \leftrightarrow \beta)\,,
\end{equation}
with $\bar{\nu}\neq \nu$. The quantum metric tensor $g^{\alpha\beta}_\nu$ measures the distance between neighbouring Bloch states, and is defined as $g^{\alpha\beta}_\nu = \sum_{ij}\mathrm{Re}[A^\alpha_{\nu i,\bar{\nu} j}A^\beta_{\bar{\nu} j,\nu i}]$, with $\bar{\nu}\neq \nu$. $v^{\alpha}_\nu g^{\alpha\beta}_\nu$ can be regarded as the quantum metric tensor dipole. For $\INH^{yxx}$, the two relevant quantum metric dipoles are $v^{y}_\nu g^{xx}_\nu$ and $v^{x}_\nu g^{yx}_\nu$, which are plotted for the lower band in Fig.~\ref{fig:Dirac}(c,d). $\INH^{yxx}$ is determined by the difference between these two dipoles, and hence nonzero as observed from the figure.

\Red{In some antiferromagnets, massless Dirac points can also be realized with certain crystalline symmetries~\cite{tang2016dirac,shao2019dirac,smejkal-ElectricControlDirac-2017}. For massless Dirac  points with $m\rightarrow 0$ in Eq.~\eqref{Diracham}, the expression of $\INH$ [Eq.~\eqref{inhexp}] diverges when the chemical potential lies exactly at the Dirac point. Disorder is the most likely source that will regularize the behavior of second order conductivity for band touchings, and further theoretical investigation is needed.}

\begin{figure}
\centering
\includegraphics[width=0.7285\columnwidth]{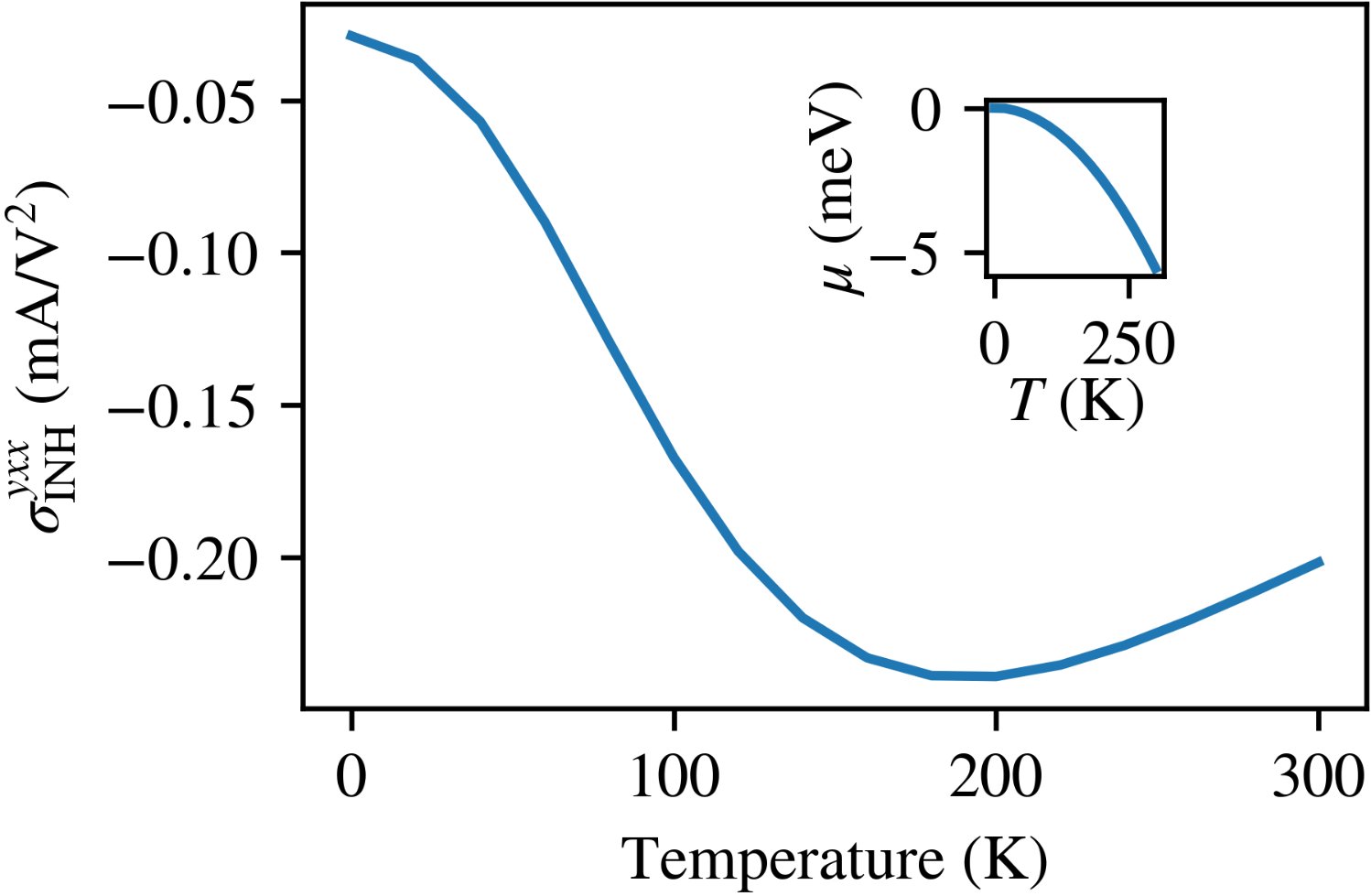}
\caption{The temperature dependence of intrinsic nonlinear Hall conductivity in tetragonal CuMnAs. Inset: temperature dependence of chemical potential calculated with the assumption of constant electron density.}
\label{fig:extra}
\end{figure}

\textit{Temperature dependence}.---By multiplying Eq.~(\ref{inhexp}) with $\int \mathd\lambda \delta(\epsilon_n - \lambda)$, the INHE at finite temperatures can be expressed as a weighted integration of the INHE at zero temperature around the chemical potential $\mu$ as $\INH(T,\mu) = \int \mathd\lambda \INH(T=0, \lambda) [-\partial f(\lambda; T, \mu)/ \partial \lambda]$. Finite temperature modifies $\INH$ in two ways: $T$ directly enters the Fermi-Dirac distribution and $\mu$ has a small dependence on temperature.

Assuming finite electron density, we calculate the dependence of chemical potential on $T$. $\mu$ decreases quadratically with respect to $T$, as expected from the Sommerfeld expansion. At room temperature, $\mu$ drops by 5~meV. Figure~\ref{fig:extra} shows the behavior of $\INH$ as a function of $T$, where both $\INH^{yxx}$ and $\INH^{yzz}$ have a strong dependence on temperature. Especially, $\INH^{yxx}$ receives an order of magnitude enhancement near room temperature. We find that this drastic change comes from the broadening of $-\partial f(\lambda; T, \mu)/ \partial \lambda$, which can take advantage of the large INHE away from the intrinsic chemical potential; the change of $\mu$ plays a negligible role in the temperature dependence of $\INH$.

\textit{Second order Drude conductivity}.---$\INH$ is not the only member of $T$-odd second order conductivity. Especially, the Drude conductivity can also be generalized to second order as~\cite{holder2020consequences,watanabe2020nonlinear,watanabe_chiral_2021,zelezny-UnidirectionalMagnetoresistanceSpinorbit-2021}
\begin{equation}
\Drude^{\alpha\beta\gamma} = - \frac{e^3 \tau^2}{\hbar^3} \sum_n \int \frac{\mathd^3 \bm{k}}{ (2\uppi)^3}(\partial_{k^\alpha}\partial_{k^\beta}\partial_{k^\gamma} \epsilon_n)f(\epsilon_n).\label{drudeexp}
\end{equation}
$\Drude^{\alpha\beta\gamma}$ is also a Fermi surface property, because an integration by parts can bring the $k$ derivative to $f(\epsilon_n)$. In contrast to the INHE, $\Drude^{\alpha\beta\gamma}$ only depends on the band dispersion and is symmetric with respect to the permutation of all its Cartesian indices. Since $\Drude$ is proportional to $\tau^2$, it also requires time reversal symmetry breaking and is allowed by $PT$ symmetry. In moderately conducting samples, $\INH$ is expected to dominate over $\Drude$. \Red{Our estimation indeed shows that $\Drude$ is only a fraction of $\INH$ for tetragonal CuMnAs at room temperature at the intrinsic Fermi energy~\cite{supplemental}}.

\textit{Magnetic point groups for INHE observation}.---Tetragonal CuMnAs is only one example where the INHE is prominent. To find other materials candidates, symmetry guidelines are needed. Symmetry groups that forbid $\BCD$ but not $\INH$ can be found based on the fact that they transform oppositely under time reversal operation. $\Drude$ can be distinguished from $\INH$ and $\BCD$ by noticing $\Drude$ transforms as a symmetric rank-3 tensor while both $\BCD$ and $\INH$ transform as rank-2 pseudo-tensors under point-group operations. For example, $C_{4z}$ constrains second order conductivity as $\sigma^{xyz} = - \sigma^{yxz}$, which is compatible with $\INH$ and $\BCD$ but not $\Drude$.

\begin{table}
\begin{tabular}{|p{5.5cm}|c|c|c|}
\hline
Magnetic point groups &
INHE &
Drude &
BCD \\
\hline
\mbox{4/m$'$m$'$m$'$}, \mbox{-6$'$m$'$2}, \mbox{6/m$'$m$'$m$'$} & \tick & \cross & \cross \\
\hline
\mbox{-6}, \mbox{6$'$/m}, \mbox{-6m2}, \mbox{-6m$'$2$'$}, \mbox{6$'$/mmm$'$}, \mbox{23}, \mbox{m$'$-3$'$}, \mbox{4$'$32$'$}, \mbox{-43m}, \mbox{m$'$-3$'$m} & \cross & \tick & \cross \\
\hline
\mbox{11$'$}, \mbox{21$'$}, \mbox{m1$'$}, \mbox{2221$'$}, \mbox{mm21$'$}, \mbox{41$'$}, \mbox{-41$'$}, \mbox{4221$'$}, \mbox{4mm1$'$}, \mbox{-42m1$'$}, \mbox{31$'$}, \mbox{321$'$}, \mbox{3m1$'$}, \mbox{61$'$}, \mbox{6221$'$}, \mbox{6mm1$'$} & \cross & \cross & \tick \\
\hline
\mbox{-1$'$}, \mbox{2$'$/m}, \mbox{2/m$'$}, \mbox{m$'$mm}, \mbox{m$'$m$'$m$'$}, \mbox{4/m$'$}, \mbox{4$'$/m$'$}, \mbox{4/m$'$mm}, \mbox{4$'$/m$'$m$'$m}, \mbox{-3$'$}, \mbox{-3$'$m}, \mbox{-3$'$m$'$}, \mbox{-6$'$}, \mbox{6/m$'$}, \mbox{-6$'$m2$'$}, \mbox{6/m$'$mm} & \tick & \tick & \cross \\
\hline
\mbox{422}, \mbox{4m$'$m$'$}, \mbox{-4$'$2m$'$}, \mbox{622}, \mbox{6m$'$m$'$} & \tick & \cross & \tick \\
\hline
\mbox{6$'$}, \mbox{6$'$22$'$}, \mbox{6$'$mm$'$} & \cross & \tick & \tick \\
\hline
\mbox{1},\mbox{2}, \mbox{2$'$}, \mbox{m}, \mbox{m$'$}, \mbox{222}, \mbox{2$'$2$'$2}, \mbox{mm2}, \mbox{m$'$m2$'$}, \mbox{m$'$m$'$2}, \mbox{4}, \mbox{4$'$}, \mbox{-4}, \mbox{-4$'$}, \mbox{4$'$22$'$}, \mbox{42$'$2$'$}, \mbox{4mm}, \mbox{4$'$m$'$m}, \mbox{-42m}, \mbox{-4$'$2$'$m}, \mbox{-42$'$m$'$}, \mbox{3}, \mbox{32}, \mbox{32$'$}, \mbox{3m}, \mbox{3m$'$}, \mbox{6}, \mbox{62$'$2$'$}, \mbox{6mm} & \tick & \tick & \tick  \\
\hline
\end{tabular}
\caption{Magnetic point groups classified by the existence or absence of INHE, second order Drude conductivity and BCD-induced NHE.}\label{table:mpgs}
\end{table}

Table~\ref{table:mpgs} presents a classification of magnetic point groups according to the existence or absence of the INHE, second order Drude conductivity and BCD contribution~\footnote{For a second order conductivity tensor $\sigma^{\alpha\beta\gamma}$, only the $\beta \leftrightarrow \gamma$ permutation symmetric part contributes to the current. Therefore, in Table~\ref{table:mpgs}, we have an extra constraint that such symmetric part does not vanish.}. There are indeed three magnetic point groups where the INHE is allowed but second order Drude conductivity and BCD contribution are forbidden (first row of Table~\ref{table:mpgs}). For the three groups, the allowed $\INH$ components are of the type $\INH^{xyz}$. More magnetic point groups can be included if we allow for the existence of second order Drude conductivity. 16 magnetic point groups belong to this class (fourth row of Table~\ref{table:mpgs}), which includes 2$'$/m, the magnetic point group of tetragonal CuMnAs. Out of all 122 magnetic point groups, 53 allows INHE (Table~\ref{table:mpgs}). Some of the groups allow both INHE and BCD contribution, enabling the comparison between these two contributions.

There are also magnetic point groups that allow BCD contribution but forbid INHE (third and sixth row of Table~\ref{table:mpgs}), most of which contain time reversal symmetry. It is worth noting that some antiferromagnetic materials actually belong to these magnetic point groups. For example, antiferromagnetic \Red{cubic} CuMnSb breaks the inversion symmetry but respects the combination of time reversal and translation symmetry, and the corresponding magnetic point group will have time reversal as a symmetry element. As a result, both the INHE and the second order Drude contribution are forbidden, but BCD contribution is allowed and can be used to detect the 90$^\circ$ \Red{(but not 180$^\circ$)} reorientation of the N\'eel vector~\cite{shao-NonlinearAnomalousHall-2020}.

\textit{Discussion and summary}.---In a recent experiment on tetragonal CuMnAs, the second order conductivity has been measured~\cite{godinho_electrically_2018}. While the dependence of the conductivity on the rotation angle of the N\'eel vector agrees with our symmetry analysis, the experimental value is at least one order of magnitude smaller than our calculation. Several factors could play a role here, including the highly sensitive doping- and temperature-dependence of the INHE and the existence of antiferromagnetic domains~\cite{janda-MagnetoSeebeckMicroscopyDomain-2020,wornle-CurrentinducedFragmentationAntiferromagnetic-2019,grzybowski-ImagingCurrentInducedSwitching-2017}. More investigations are needed for a quantitative comparison between our theory and the experiment.

Beyond isotropic relaxation approximation, skew-scattering and side-jump arising from disorder effects also contribute to the second order conductivity, but INHE remains the only contribution that is independent of the relaxation time. Extensive efforts~\cite{isobe_high-frequency_2020,du2019disorder,nandy2019symmetry,xiao2019theory,du2020quantum,konig2021quantum,Du.2021.Xie} have been put into disorder-induced NHE in time reversal symmetric systems. The behavior of these contributions in magnetic systems and their role in possible spintronics applications remains to be explored.

In summary, we have shown that the INHE leads to a dominant contribution to the second order conductivity in tetragonal CuMnAs. More importantly, we find that the INHE can not only reflect the microscopic geometric properties of Bloch electrons, but also respond sensitively to the orientation of the N\'eel vector, providing a promising way to detect the flipping of the N\'eel vector. Our symmetry analysis shows that the INHE is widely available and hence should be recognized as a fundamental transport phenomenon in antiferromagnetic spintronics.

\begin{acknowledgments}
\textit{Acknowledgments}.---We acknowledge stimulating discussions with T.~Jungwirth, D.C.~Ralph, and J.~{\v Z}elezn{\'y}. This work is supported by AFOSR MURI 2D MAGIC (FA9550-19-1-0390). Y.~G. also acknowledges support from the Startup Foundation from USTC. This work used the Extreme Science and Engineering Discovery Environment (XSEDE), which is supported by National Science Foundation grant number ACI-1548562. Specifically, it used the Bridges-2 system, which is supported by NSF award number ACI-1928147, at the Pittsburgh Supercomputing Center (PSC).
\end{acknowledgments}

\bibliography{main}

\end{document}


\title{Supplemental Material for ``Intrinsic nonlinear Hall effect in antiferromagnetic tetragonal CuMnAs''}

\author{Chong \surname{Wang}}
\affiliation{Department of Physics, Carnegie Mellon University, Pittsburgh, Pennsylvania 15213, USA}

\author{Yang \surname{Gao}}
\email{ygao87@ustc.edu.cn}
\affiliation{ICQD, Hefei National Laboratory for Physical Sciences at Microscale, University of Science and Technology of China, Hefei, Anhui 230026, China}
\affiliation{Department of Physics, University of Science and Technology of China, Hefei, Anhui 230026, China}

\author{Di \surname{Xiao}}
\email{dixiao@cmu.edu}
\affiliation{Department of Physics, Carnegie Mellon University, Pittsburgh, Pennsylvania 15213, USA}

\maketitle

\section{Details of DFT calculations}

DFT calculations were carried out to evaluate Eq.~(1) in the main text. The \textsc{vasp} package~\cite{vasp1,vasp2} combined with the \textsc{wannier90}~\cite{pizzi2020wannier90} package was used to construct a tight binding Hamiltonian for CuMnAs. Projector augmented wave method~\cite{blochl_projector_1994} and Perdew-Burke-Ernzerhof exchange-correlation functional~\cite{perdew_generalized_1996} were utilized in the DFT calculations. A $16 \times 16 \times 10$ mesh was used to sample the Brillouin zone and the plane wave energy cutoff was 400~eV. A Hubbard $U$ of 1.7~eV~\cite{veis_band_2018} was added to account for the electron-electron repulsion on the Mn $d$ orbitals. Experimental lattice constants were used and the internal coordinates of atoms were relaxed until the forces on each atom were smaller than 1~meV/\AA.

\section{$\INH^{yzz}$ of tetragonal C\lowercase{u}M\lowercase{n}A\lowercase{s}}

\begin{figure}[h]
\centering
\includegraphics[width=0.1875\textwidth]{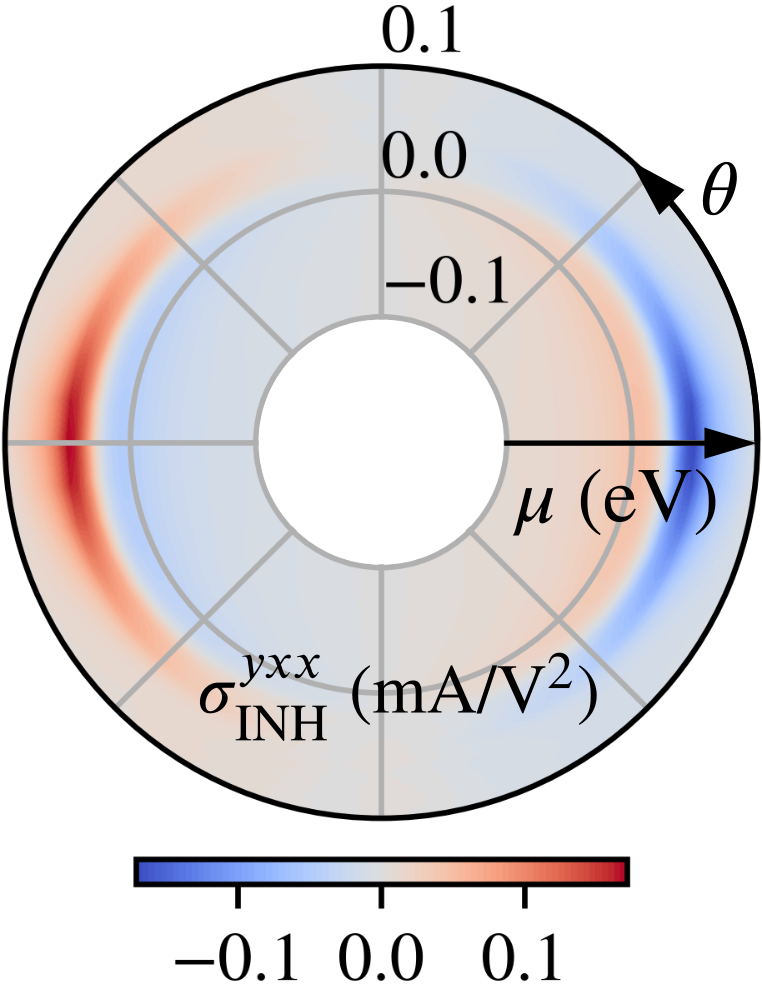}
\caption{$\INH^{yzz}$ of tetragonal CuMnAs as a function of chemical potential $\mu$ and orientation of the N\'eel vector $\theta$.}
\label{fig:yzz}
\end{figure}

As explained in the main text, the only independent components of $\INH$ are $\INH^{yxx}$ and $\INH^{yzz}$. $\INH^{yxx}$ has been extensively discussed in the main text and we present the DFT result of $\INH^{yzz}$ in Fig.~\ref{fig:yzz}.

\section{Second order Drude conductivity in C\lowercase{u}M\lowercase{n}A\lowercase{s}}

\begin{figure}
\centering
\includegraphics[width=0.5823809523809523\textwidth]{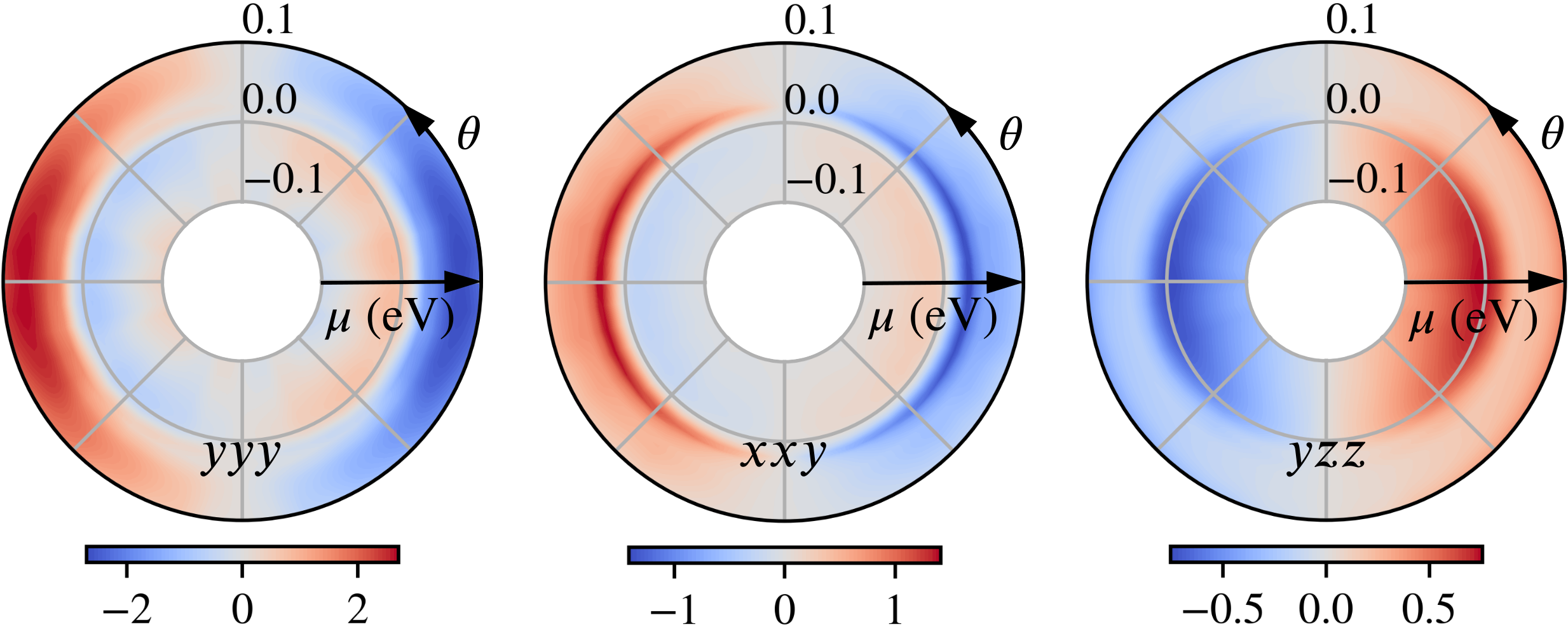}
\caption{Second order Drude weight of CuMnAs. The unit is $\mu$A$\cdot$eV$^2$/V$^2$.}
\label{fig:Drude}
\end{figure}

The expression of second order Drude conductivity has been provided in Eq.~(4) in the main text. Here we extract the $\tau$-independent part of the $\Drude$ and call it second order Drude weight: $\Drude^{\alpha\beta\gamma}=D^{\alpha\beta\gamma}/\eta^2$, with $\eta = \tau/\hbar$. With the calculation method detailed in the main text, we find that the second order Drude weight is on the order of $\mu$A$\cdot$eV$^2$/V$^2$ (Fig.~\ref{fig:Drude}). To estimate the value of $\eta$, we notice that the experimental value of resistivity is roughly $1.6 \times 10^{-4}$~$\Omega\cdot$cm at 230~K. Within relaxation approximation, the conductivity is 
\begin{equation}
  \sigma_{\mathrm{Drude}}^{\alpha \beta} =  \frac{D^{\alpha \beta}}{\eta}, 
\end{equation}
where
\begin{equation}
  D^{\alpha \beta}  =  - e^2 \hbar \sum_n \int_{\mathrm{BZ}} \frac{\mathd
  \bm{k}}{(2 \pi)^3} \frac{\partial f_{n\bm{k}}}{\partial
  \epsilon_{n\bm{k}}} v^{\alpha}_{n \bm{k}}
  v_{n\bm{k}}^{\beta} 
\end{equation}
is the (linear) Drude weight. In our DFT calculations we find $D^{xx}$ and $D^{yy}$ are roughly the same, with the value 882~eV$\cdot$($\Omega\cdot$cm)$^{-1}$. Therefore, $\eta$ is estimated to be 0.16~eV. The largest component of second order Drude weight at intrinsic Fermi energy for $\theta=0$ is $D^{xyy}=0.75$~$\mu$A$\cdot$eV$^2$/V$^2$, corresponding to $\Drude^{xyy}=0.029$~mA/V$^2$, which is much smaller than $\INH^{yxx}\approx 0.2$~mA/V$^2$ at room temperature (Fig.~4 in the main text).

\section{Intrinsic nonlinear Hall conductivity and geometric tensor}

The quantum metric tensor measures the distance between neighbouring Bloch
state:
\begin{eqnarray}
  \mathrm{Tr} (P_{\nu \bm{k}}) - \mathrm{Tr} (P_{\nu \bm{k}} P_{\nu
  \bm{k}+ \delta \bm{k}}) & \approx & g^{\alpha \beta}_{\nu}
  (\bm{k}) \delta k^{\alpha} \delta k^{\beta},
\end{eqnarray}
where repeated Cartesian indices imply summation and $P_{\nu \bm{k}}$ is
the projection operator
\begin{eqnarray}
  P_{\nu \bm{k}} & = & \sum_{n \in \nu} \ket{u_{n 
  \bm{k}}} \bra{u_{n  \bm{k}}} .
\end{eqnarray}
$u$ is the periodic part of the Bloch state. $\nu$ denotes a band set. If a
band is nondegenerate, $\nu$ only contains this band. If there are degenerate
bands, $\nu$ contains all these bands. Written explicitly,
\begin{eqnarray}
  g^{\alpha \beta}_{\nu} (\bm{k}) \delta k^{\alpha} \delta k^{\beta} & =
  & \sum_{n, m \in \nu} \delta_{n  m} - \left|
  \braket{u_{n\bm{k}}}{u_{m\bm{k}+ \delta \bm{k}}} \right|^2
  .
\end{eqnarray}
To evaluate $g (\bm{k})$, we expand $\delta_{n  m} - \left|
\braket{u_{n\bm{k}}}{u_{m\bm{k}+ \delta \bm{k}}} \right|^2$,
where $n$ and $m$ are band indices. We first expand
$\braket{u_{n\bm{k}}}{u_{m\bm{k}+ \delta \bm{k}}}$ to second
order as
\begin{eqnarray}
  \braket{u_{n\bm{k}}}{u_{m\bm{k}+ \delta \bm{k}}} & \approx
  & \braket{u_{n\bm{k}}}{\partial_{\alpha} u_{m\bm{k}}} \delta
  k^{\alpha} + \frac{1}{2} \braket{u_{n\bm{k}}}{\partial_{\alpha}
  \partial_{\beta} u_{m\bm{k}}} \delta k^{\alpha} \delta k^{\beta} +
  \delta_{n  m}
\end{eqnarray}
and then collect terms of different orders from $\delta_{n  m} - \left|
\braket{u_{n\bm{k}}}{u_{m\bm{k}+ \delta \bm{k}}} \right|^2$.
The zeroth order term in $\delta \bm{k}$ is clearly zero. Since
$\delta_{n  m} - \left| \braket{u_{n\bm{k}}}{u_{m\bm{k}+
\delta \bm{k}}} \right|^2$ reaches maximum or minimum at $\delta
\bm{k}=\bm{0}$, the linear term should also be zero. We verify
this statement by writing out the linear terms:
\begin{eqnarray}
  &  & - \delta_{n  m} \delta k^{\alpha} \left(
  \braket{u_{n\bm{k}}}{\partial_{\alpha} u_{m\bm{k}}} +
  \braket{\partial_{\alpha} u_{m\bm{k}}}{u_{n\bm{k}}} \right)\\
  & = & - \delta_{n  m} \delta k^{\alpha} \partial_{\alpha}
  \braket{u_{n\bm{k}}}{u_{m\bm{k}}}\\
  & = & - \delta_{n  m} \delta k^{\alpha} \partial_{\alpha} \delta_{n
   m}\\
  & = & 0,
\end{eqnarray}
where $\partial_{\alpha} \equiv \partial_{k^{\alpha}}$. The second order
terms are
\begin{eqnarray}
  &  & - \left( \braket{u_{n\bm{k}}}{\partial_{\alpha}
  u_{m\bm{k}}} \braket{\partial_{\beta}
  u_{m\bm{k}}}{u_{n\bm{k}}} + \frac{1}{2}
  \braket{u_{n\bm{k}}}{\partial_{\alpha} \partial_{\beta}
  u_{m\bm{k}}} \delta_{n  m} + \frac{1}{2}
  \braket{\partial_{\alpha} \partial_{\beta}
  u_{m\bm{k}}}{u_{n\bm{k}}} \delta_{n  m} \right) \delta
  k^{\alpha} \delta k^{\beta} .
\end{eqnarray}
To simplify the expression, we use the following identities
\begin{eqnarray}
  \braket{u_{n\bm{k}}}{u_{m\bm{k}}} & = & \delta_{n  m},\\
  \braket{u_{n\bm{k}}}{\partial_{\alpha} u_{m\bm{k}}} +
  \braket{\partial_{\alpha} u_{n\bm{k}}}{u_{m\bm{k}}} & = & 0,\\
  \braket{\partial_{\beta} u_{n\bm{k}}}{\partial_{\alpha}
  u_{m\bm{k}}} + \braket{u_{n\bm{k}}}{\partial_{\alpha}
  \partial_{\beta} u_{m\bm{k}}} + \braket{\partial_{\alpha}
  \partial_{\beta} u_{n\bm{k}}}{u_{m\bm{k}}} +
  \braket{\partial_{\alpha} u_{n\bm{k}}}{\partial_{\beta}
  u_{m\bm{k}}} & = & 0,
\end{eqnarray}
where each identity is the derivative of the previous one. Therefore,
\begin{eqnarray}
  &  & - \left( \braket{u_{n\bm{k}}}{\partial_{\alpha}
  u_{m\bm{k}}} \braket{\partial_{\beta}
  u_{m\bm{k}}}{u_{n\bm{k}}} + \frac{1}{2}
  \braket{u_{n\bm{k}}}{\partial_{\alpha} \partial_{\beta}
  u_{m\bm{k}}} \delta_{n  m} + \frac{1}{2}
  \braket{\partial_{\alpha} \partial_{\beta}
  u_{m\bm{k}}}{u_{n\bm{k}}} \delta_{n  m} \right) \delta
  k^{\alpha} \delta k^{\beta}\\
  & = & \left( - A_{n  m}^{\alpha} A_{n  m}^{\beta} +
  \frac{1}{2} \braket{\partial_{\beta} u_{n\bm{k}}}{\partial_{\alpha}
  u_{m\bm{k}}} \delta_{n  m} + \frac{1}{2}
  \braket{\partial_{\alpha} u_{n\bm{k}}}{\partial_{\beta}
  u_{m\bm{k}}} \delta_{n  m} \right) \delta k^{\alpha} \delta
  k^{\beta}\\
  & = & \left( - A_{n  m}^{\alpha} A_{n  m}^{\beta} +
  \frac{1}{2} \sum_l A^{\beta}_{n  l} A^{\alpha}_{l  n}
  \delta_{n  m} + \frac{1}{2} \sum_l A^{\alpha}_{n  l}
  A^{\beta}_{l  n} \delta_{n  m} \right) \delta k^{\alpha}
  \delta k^{\beta},
\end{eqnarray}
where $A_{n  m}^{\alpha} = \mathrm{i}
\braket{u_{n\bm{k}}}{\partial_{\alpha} u_{m\bm{k}}}$ is Berry
connection. Therefore
\begin{eqnarray}
  g^{\alpha \beta}_{\nu} (\bm{k}) & = & \sum_{n, m \in \nu} \left(
  \frac{1}{2} \sum_l A^{\beta}_{n  l} A^{\alpha}_{l  n}
  \delta_{n  m} + \frac{1}{2} \sum_l A^{\alpha}_{n  l}
  A^{\beta}_{l  n} \delta_{n  m} - A_{n  m}^{\alpha}
  A_{n  m}^{\beta} \right)\\
  & = & \sum_{n \in \nu, l \not{\in} \nu} \mathrm{Re} (A^{\alpha}_{nl}
  A^{\beta}_{ln}) .
\end{eqnarray}
The intrinsic nonlinear Hall conductivity is
\begin{eqnarray}
  \sigma^{\alpha \beta \gamma}_{\mathrm{INH}} & = & 2 e^3 \sum_{n,
  m}^{\epsilon_m \neq \epsilon_n} \mathrm{Re} \int \frac{\mathd
  \bm{k}}{(2 \uppi)^3} \frac{v^{\alpha}_n A^{\beta}_{n  m}
  A^{\gamma}_{m  n}}{\epsilon_n - \epsilon_m} \frac{\partial f
  (\epsilon_n)}{\partial \epsilon_n} - (\alpha \leftrightarrow \beta) .
\end{eqnarray}
For a system with two band sets $\nu \in \{ 1, 2 \}$, where within each band
set $\nu$ there are two degenerate bands $i \in \{ 1, 2 \}$,
\begin{eqnarray}
  \sigma^{\alpha \beta \gamma}_{\mathrm{INH}} & = & 2 e^3 \sum_{\nu} \sum_{i \in
  \nu} \sum_{j \in \bar{\nu}} \mathrm{Re} \int \frac{\mathd \bm{k}}{(2
  \uppi)^3} \frac{v^{\alpha}_{\nu} A^{\beta}_{i  j} A^{\gamma}_{j
   i}}{\epsilon_{\nu} - \epsilon_{\bar{\nu}}} \frac{\partial f
  (\epsilon_n)}{\partial \epsilon_n} - (\alpha \leftrightarrow \beta)\\
  & = & 2 e^3 \sum_{\nu} \int \frac{\mathd \bm{k}}{(2
  \uppi)^3} \frac{v^{\alpha}_{\nu} g_{\nu}^{\beta \gamma}}{\epsilon_{\nu}
  - \epsilon_{\bar{\nu}}} \frac{\partial f (\epsilon_n)}{\partial
  \epsilon_n} - (\alpha \leftrightarrow \beta) .
\end{eqnarray}
This is Eq.~(3) in the main text.

\bibliography{main}